\documentclass[10pt,aps,prx,twocolumn,superscriptaddress,nofootinbib]{revtex4-2}
\usepackage[utf8]{inputenc}
\usepackage[T1]{fontenc}
\usepackage{braket}
\usepackage{amsfonts, mathtools}
\usepackage{dcolumn}
\usepackage{booktabs}
\usepackage{siunitx}
\usepackage[breaklinks=true,colorlinks,citecolor=blue,linkcolor=blue,urlcolor=blue]{hyperref}
\usepackage{comment}
\usepackage{orcidlink}
\usepackage[ruled,vlined]{algorithm2e}
\usepackage{xcolor}
\usepackage{float}

\usepackage{listings}
\lstdefinestyle{kipu}{
  basicstyle=\ttfamily\small,
  numbers=left,
  numberstyle=\tiny,
  stepnumber=1,
  numbersep=8pt,
  showstringspaces=false,
  breaklines=true,
  frame=single,
  rulecolor=\color{black!20},
  tabsize=2,
  captionpos=b,
  upquote=true
}
\lstdefinelanguage{json}{
  basicstyle=\ttfamily\small,
  literate=
   *{0}{{{\color{black}0}}}{1}
    {1}{{{\color{black}1}}}{1}
    {2}{{{\color{black}2}}}{1}
    {3}{{{\color{black}3}}}{1}
    {4}{{{\color{black}4}}}{1}
    {5}{{{\color{black}5}}}{1}
    {6}{{{\color{black}6}}}{1}
    {7}{{{\color{black}7}}}{1}
    {8}{{{\color{black}8}}}{1}
    {9}{{{\color{black}9}}}{1}
    {:}{{{\color{black}{:}}}}{1}
    {,}{{{\color{black}{,}}}}{1}
    {"}{{{\color{black}{"}}}}{1},
}

\usepackage{soul}

\begin{document}

\title{Digitized Counterdiabatic Quantum Feature Extraction}

\author{Anton Simen$^{\orcidlink{0000-0001-8863-4806}}$}
\email{anton.simen@kipu-quantum.com}
\affiliation{Kipu Quantum GmbH, Greifswalderstrasse 212, 10405 Berlin, Germany}
\affiliation{Department of Physical Chemistry, University of the Basque Country UPV/EHU, Apartado 644, 48080 Bilbao, Spain}

\author{Carlos Flores-Garrig\'os$^{\orcidlink{0009-0000-9735-5411}}$}
\affiliation{Kipu Quantum GmbH, Greifswalderstrasse 212, 10405 Berlin, Germany}
\affiliation{IDAL, Electronic Engineering Department, ETSE-UV, University of Valencia, Avgda. Universitat s/n, 46100 Burjassot, Valencia, Spain}

\author{Murilo Henrique De Oliveira$^{\orcidlink{0000-0001-9777-8342}}$}
\affiliation{Kipu Quantum GmbH, Greifswalderstrasse 212, 10405 Berlin, Germany}

\author{Gabriel Dario Alvarado Barrios$^{\orcidlink{0000-0002-8684-4209}}$}
\affiliation{Kipu Quantum GmbH, Greifswalderstrasse 212, 10405 Berlin, Germany}

\author{Alejandro Gomez Cadavid$^{\orcidlink{0000-0001-8863-4806}}$}
\affiliation{Kipu Quantum GmbH, Greifswalderstrasse 212, 10405 Berlin, Germany}
\affiliation{Department of Physical Chemistry, University of the Basque Country UPV/EHU, Apartado 644, 48080 Bilbao, Spain}

\author{Archismita Dalal$^{\orcidlink{0000-0003-0638-8328}}$}
\email{archismita.dalal@kipu-quantum.com}
\affiliation{Kipu Quantum GmbH, Greifswalderstrasse 212, 10405 Berlin, Germany}

\author{Enrique Solano$^{\orcidlink{0000-0002-8602-1181}}$}
\email{enr.solano@gmail.com}
\affiliation{Kipu Quantum GmbH, Greifswalderstrasse 212, 10405 Berlin, Germany}

\author{Narendra N. Hegade$^{\orcidlink{0000-0002-9673-2833}}$}
\affiliation{Kipu Quantum GmbH, Greifswalderstrasse 212, 10405 Berlin, Germany}

\author{Qi Zhang$^{\orcidlink{0000-0001-6223-5516}}$}
\email{qizhang.zm@gmail.com}
\affiliation{Kipu Quantum GmbH, Greifswalderstrasse 212, 10405 Berlin, Germany}

\date{\today}

\begin{abstract}
We introduce a Hamiltonian-based quantum feature extraction method that generates complex features via the dynamics of $k$-local many-body spins Hamiltonians, enhancing machine learning performance. Classical feature vectors are embedded into spin-glass Hamiltonians, where both single-variable contributions and higher-order correlations are represented through many-body interactions. By evolving the system under suitable quantum dynamics on IBM digital quantum processors with 156 qubits, the data are mapped into a higher-dimensional feature space via expectation values of low- and higher-order observables. This allows us to capture statistical dependencies that are difficult to access with standard classical methods. We assess the approach on high-dimensional, real-world datasets, including molecular toxicity classification and image recognition, and analyze feature importance to show that quantum-extracted features complement and, in many cases, surpass classical ones. The results suggest that combining quantum and classical feature extraction can provide consistent improvements across diverse machine learning tasks, indicating a reliable level of early quantum usefulness for near-term quantum devices in data-driven applications.
\end{abstract}

\maketitle

\section{Introduction}\label{sec:introduction}

Feature extraction is a key step in machine learning, where raw data is transformed into informative representations that enable models to recognize patterns, improve accuracy, and generalize effectively. Quantum systems offer a natural mechanism for generating rich features since their dynamics and correlations can encode non-linear relationships that are difficult to capture with classical preprocessing methods.

Quantum machine learning (QML) leverages this potential through several complementary approaches. Quantum feature maps embed classical data into high-dimensional Hilbert spaces, allowing classical algorithms to operate in more expressive representations~\cite{nkilloran2019, havlivcek2019supervised,simen2025quenchedquantumfeaturemaps,simen2024digital}. Quantum reservoir computing (QRC), in contrast, exploits the intrinsic dynamics of many-body quantum systems as a fixed, non-linear transformation that generates expressive temporal and static features \cite{fujii2016, fujii2020review, kornjača2024largescalequantumreservoirlearning, reservoir2025quantum, beaulieu2024robust}. Both strategies highlight how quantum evolutions can serve as a powerful tool for enhancing machine learning tasks.

In this work, we propose a digital quantum feature extraction framework inspired by these principles. Using IBM's 156-qubit digital quantum processor, \textit{ibm\_kingston}, we encode the classical feature vectors into many-body $k$-local spin-glass Hamiltonians, where both single-variable terms and higher-order correlations are represented via spin interactions. Then, we drive the system through a controlled approximate counterdiabatic dynamics, specifically in the impulse regime as previously studied in the context of combinatorial optimization~\cite{sels2017, Carolan_2022, Cadavid_2024}. This evolution partially suppresses non-adiabatic transitions, which enriches the resulting quantum state while allowing for non-linear mixing of input correlations. Next, we measure the expectation values of low- and high-order observables from the resulting state to construct feature maps that capture complex statistical dependencies.

We evaluate the framework on two representative problems: molecular toxicity prediction~\cite{toxicity2021} and breast tumor detection using an image-based dataset~\cite{medmnistv2}. In both cases, quantum-derived features consistently enhance the performance of classical models when combined with standard preprocessing techniques. These results suggest that integrating Hamiltonian-based encoding with quantum dynamics provides a scalable and effective route to obtain expressive feature representations on near-term quantum devices.

\section{Hamiltonian-based encoding}
Previous works have primarily employed encoding schemes where classical feature vectors are mapped onto the longitudinal fields of spin-glass Hamiltonians \cite{simen2025quenchedquantumfeaturemaps, kornjača2024largescalequantumreservoirlearning}. Here, we extend this idea by incorporating correlation-aware encoding, in which statistical dependencies between classical variables are embedded into higher-order coupling terms. This design enables the extraction of richer and more statistically dependent features from the training dataset, denoted as $\mathbf{X}$. The corresponding Hamiltonian that encodes both individual features and their higher-order correlations is expressed as

\begin{equation}\label{eq:encoding_hamiltonian}
  H(\mathbf{x})
  = \sum_{i=1}^n x_i \,\sigma_i^z
  + \sum_{k=2}^{K} 
      \sum_{S \in \mathcal{G}^{(k)}}
      c_{S}\, \prod_{i\in S} \sigma_{i}^z ,
\end{equation}
where $x_i$ is the $i$-th component of the input feature vector $\mathbf{x}$, with $\mathbf{x} \in \mathbf{X}$. The coefficients $c_S$ capture higher-order mutual information between the variables indexed by the subset $S$ and are estimated from the dataset’s joint statistics (see Appendix~\ref{app:c}). The hypergraph $\mathcal{G}^{(k)}$ defines the structure of the $k$-body interaction terms, which can be efficiently embedded into the native coupling map of the targeted quantum processor.

This encoding scheme provides a unified framework that captures both local and collective patterns in the data. To assess its versatility, we explore configurations with varying interaction orders and investigate how they can be constructively combined. In particular, we employ circuits based on two-body Hamiltonians from which up to three-body correlations are extracted. We also examine Hamiltonians that explicitly incorporate three-body coupling terms. These complementary configurations enrich the representational structure of the resulting quantum features.

\section{Quantum-enhanced feature extraction}

The quantum feature extraction proposed in this work relies on applying quantum dynamics where the Hamiltonian in Eq.~(\ref{eq:encoding_hamiltonian}) can be encoded. Although this work can be extended to numerous quantum systems, for the sake of illustration, we focus on the dynamics simulated through counterdiabatic protocols. Specifically, the encoding Hamiltonian is used to build an adiabatic Hamiltonian, $H_{\text{ad}}\left(t;\mathbf{x}\right) = A(t)H_i + B(t)H(\mathbf{x})$, which is then used to build the adiabatic gauge potential with a first-order nested commutator~\cite{PhysRevLett.123.090602}, given by
\begin{equation}
    \mathcal{A}(t;\mathbf{x}) = i\alpha\left[H_{\text{ad}}(t;\mathbf{x}), \partial_t H_{\text{ad}}(t;\mathbf{x})\right]
    \label{eq:agp} ,
\end{equation}
where $\alpha$ can be computed analytically for arbitrary spin-glass problems \cite{Cadavid_2025}. The trotterized quantum dynamics in then given by the time evolution operator $\mathcal{U}_T(\textbf{x})=\prod_{j=1}^{N_\text{steps}} \exp\left(-i\Delta t\left(H_\text{ad}(j\Delta t;\mathbf{x}) + \mathcal{A}(j\Delta t; \mathbf{x}) \right) \right),$
where $N_\text{steps}$ is the number of Trotter steps and $\Delta t$ is the time step size, overall simulating an evolution time $T=N_\text{steps} \Delta t$. A fast quantum dynamics aided by the first-order counterdiabatic term can be simulated with a single Trotter step of Eq. \ref{eq:agp}, since the contribution of $H_\text{ad}$ is negligible in the impulse regime \cite{Carolan_2022, Cadavid_2024}. Even though the objective here is not to find the lowest energy states, the resulting circuits restrict the Hilbert space to lower energy states, enabling higher fidelity when measuring expectation values compared to other non-restrictive ansatz. Furthermore, the unavoidable non-adiabatic transitions enabled by this fast approximate counterdiabatic driving allows for the extraction of complex patterns from the information encoded into the target Hamiltonian. 

The features are mapped to expectation values estimated from the measurements in the computational basis. We propose in this work a combination of single- and higher-body expectations to build up the feature maps. This combination enables the capture of complex quantum correlations inherent in the system that encodes classical information. Thus, the mapping from a feature vector, $\mathbf{x}$, to a higher-dimensional feature map, $\tilde{\mathbf{x}}$, reads

\begin{equation}
  \tilde{\mathbf{x}}
  = \sum_{i=1}^{n} \langle \sigma_i^z \rangle\, \mathbf{e}_i
  + \sum_{k=2}^{K} 
      \sum_{S \in \mathcal{G}^{(k)}}
        \left\langle \prod_{i \in S} \sigma_i^z \right\rangle\, \mathbf{e}_S,
  \label{eq:featuremap}
\end{equation}
where expectation values of observables are computed by sampling eigenstates from a final state. Here, $\mathbf{e}$ simply represents a basis.

\section{Methodology}

To evaluate the effectiveness of the proposed quantum feature extraction approach, we address two distinct classification problems: (i) breast tumor detection using the MedMNIST database \cite{medmnistv2}, which contains 702 samples including both training and test sets of $224\times224$ ultrasound images, and (ii) molecular toxicity classification \cite{toxicity2021}, consisting of 171 molecular samples described by 156 variables.
Since these problems differ in structure—images versus tabular data—as well as in sample size and computational demands, we adopt two tailored methodologies while preserving the same core idea. The following subsections describe each case in detail.

\subsection{Molecular toxicity classification}

\begin{figure}
\centering
\includegraphics[width=0.45\textwidth]{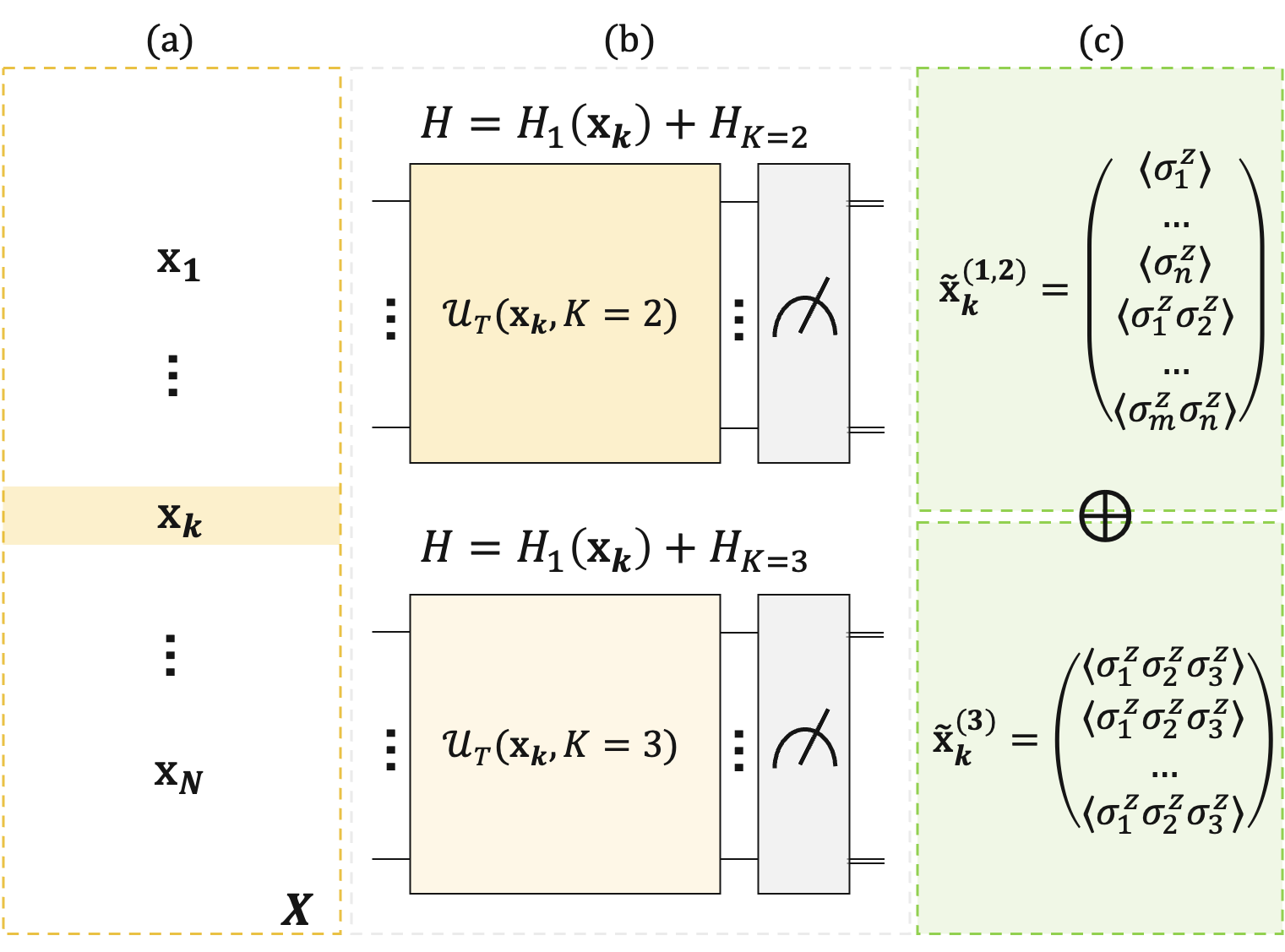}
\caption{Illustration of quantum feature extraction using multiple quantum dynamics for the molecular toxicity case. (a) Tabular dataset $X$, from which samples and classical correlations are selected. (b) The extracted information from $X$ is encoded into the local fields and higher-order coefficients of spin Hamiltonians (see Eq. \ref{eq:encoding_hamiltonian}), differing in the order of interaction terms. (c) Local magnetizations and quantum correlations are measured from the two circuits and concatenated, yielding the quantum-extracted features.}
\label{fig:schem_tox}
\end{figure}

This dataset consists of 171 molecules, each represented by 156 molecular descriptors and a binary toxicity label, toxic or non-toxic, for a specific drug design task \cite{toxicity2021}. Although the dataset is high-dimensional, its small sample size allows for more experimental flexibility with limited hardware usage. To capture richer feature representations, we apply two distinct quantum dynamics to each sample, thereby extracting complementary sets of complex features. The schematic diagram of the approach is shown in Fig. \ref{fig:schem_tox}. We employ two types of interaction terms: one involving only two-body interactions ($K=2$) and another involving only three-body interactions ($K=3$), both defined according to Eq. \ref{eq:encoding_hamiltonian}.

Feature extraction is performed by measuring one- and two-body expectation values from circuits with two-body interactions, and three-body expectation values from circuits with three-body interactions. In both cases, expectation values are computed within a closed linear chain to maintain the same feature dimensionality as the original dataset and to minimize computational overhead.

Machine learning performance is assessed using repeated stratified cross-validation with five repetitions and five splits (25 distinct test sets in total), providing a robust statistical evaluation of generalization across data partitions. The classifier employed is Gradient Boosting, with the number of estimators fixed at 1000 and a random seed of 42. 

\subsection{Breast tumor detection}
\begin{figure}
\centering
\includegraphics[width=0.45\textwidth]{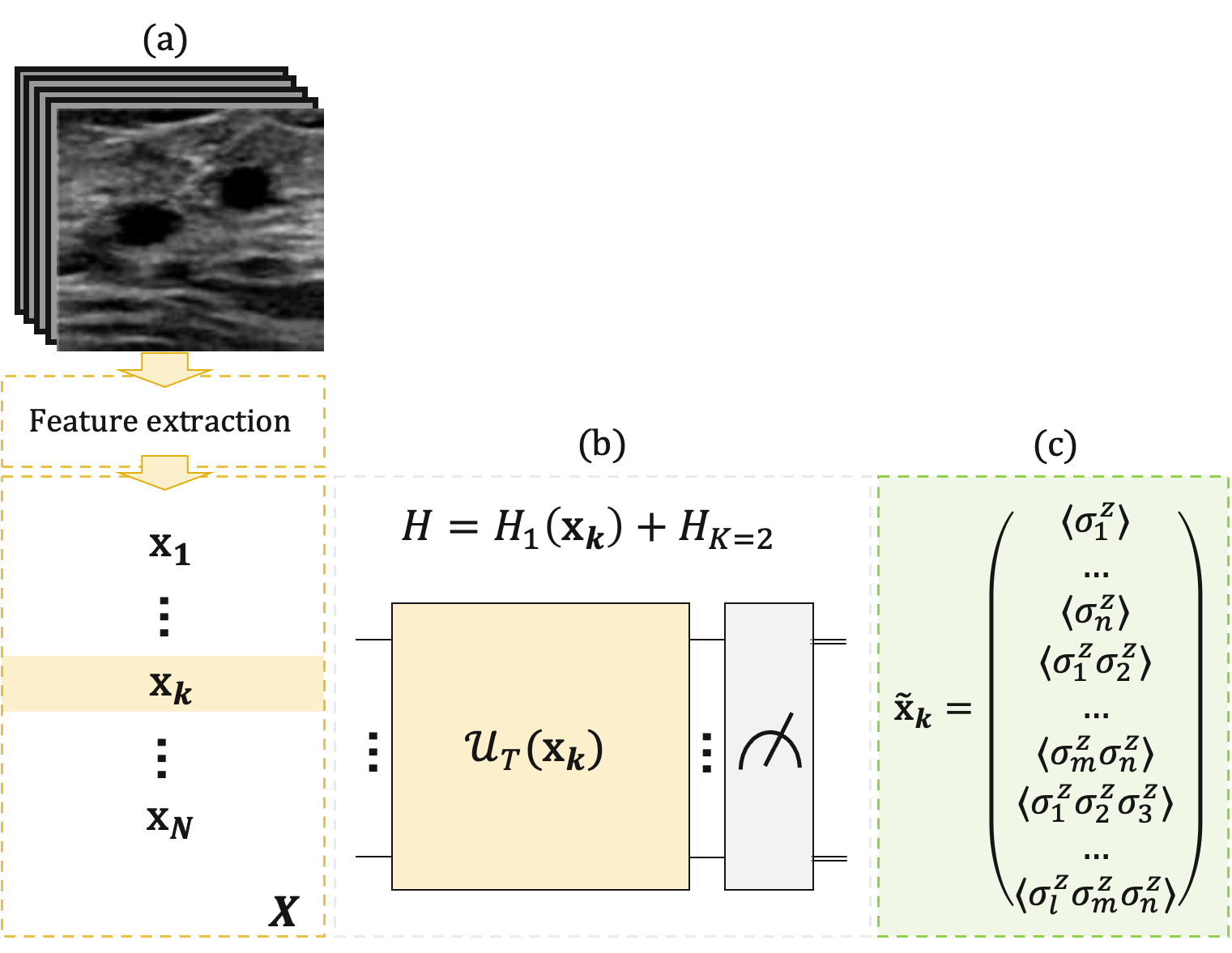}
\caption{Protocol for the image classification case, combining classical and quantum feature extraction methods. (a) Image-based dataset, where conventional feature extraction techniques are used to construct a tabular dataset $X$, from which samples and classical correlations are selected (as in Fig. \ref{fig:schem_tox}). (b) The extracted information from $X$ is encoded into the local fields and two-body coefficients of a spin Hamiltonian (see Eq. \ref{eq:encoding_hamiltonian}). (c) Local magnetizations and quantum correlations are measured and concatenated, forming the quantum-extracted features.}
\label{fig:schem_breast}
\end{figure}

For the breast tumor detection problem based on ultrasound images, we first apply classical feature extraction methods—FFT, Gabor filters, and texture-based techniques—to transform the $224\times224$ images into a tabular dataset with 202 variables. We then perform a mutual information analysis between the extracted features and target labels (using only the training set) to reduce the dimensionality to 156 variables, matching the available quantum hardware resources. This step could be extended to include feature extraction via deep learning methods \cite{featureextractioncnn2024}. The overall approach is illustrated in Fig. \ref{fig:schem_breast}.
In this case, the encoding Hamiltonian includes up to two-body interactions; however, expectation values are extracted up to the three-body level from the probability distributions in the computational basis. As in the molecular toxicity case, correlations are evaluated along a linear chain to preserve dimensionality and reduce computational cost.
Model performance is evaluated using stratified cross-validation with five splits and five repetitions, applied to all 702 samples, employing the same Gradient Boosting configuration. In addition, since this dataset has a predefined train-test split used in the MedMNIST benchmark \cite{medmnistv2}, we also report results obtained under that original configuration, as presented in Table \ref{tab:breast_test}, where results with a Support Vector Classifier with a Gaussian kernel are also presented.

\section{Results and discussion}
\begin{figure*}[!htbp] 
  \centering
  \includegraphics[width=\textwidth,height=\textheight,keepaspectratio]{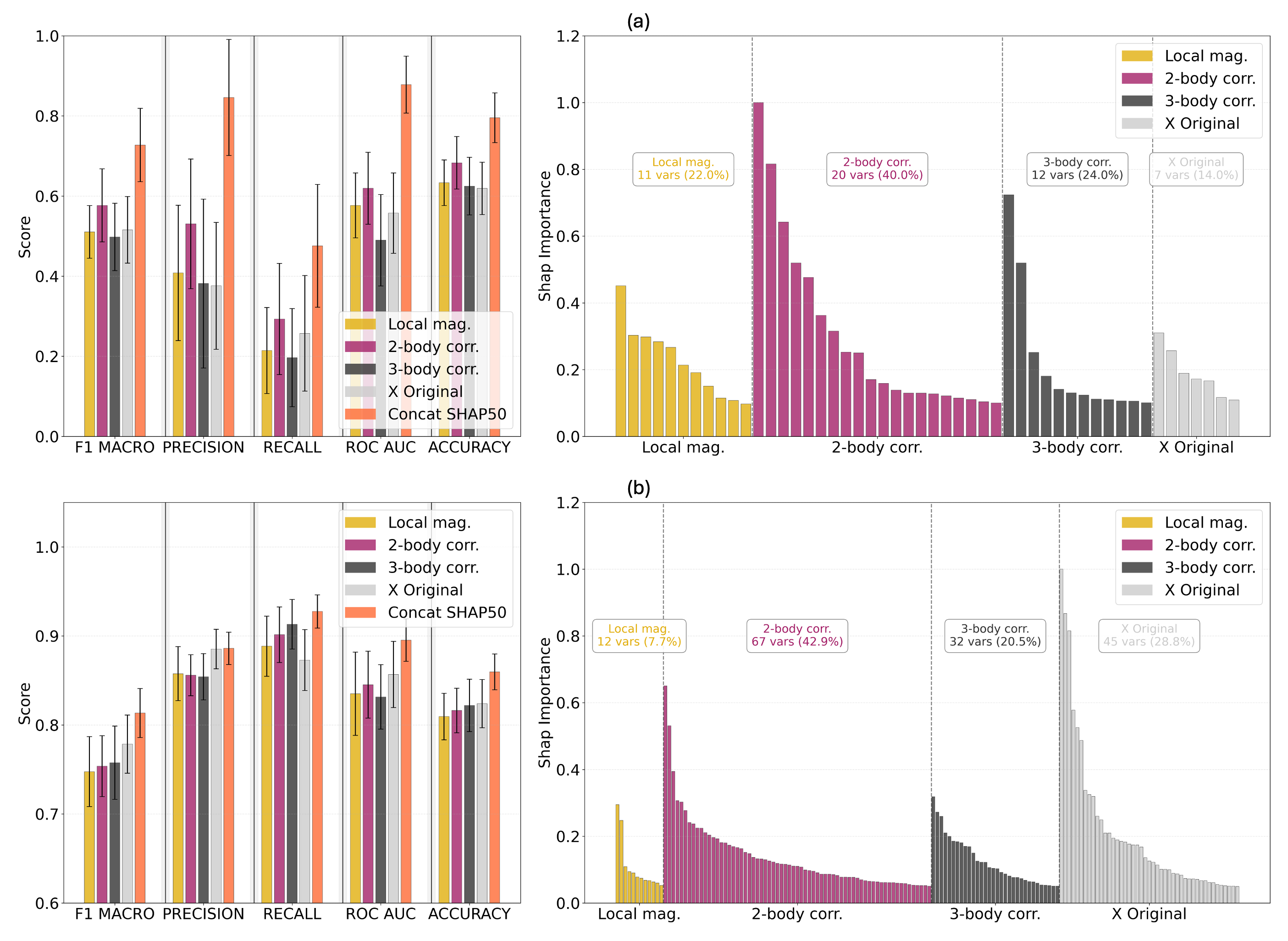}
  \caption{Performance and feature-importance analysis for (a) molecular toxicity classification and (b) breast tumor detection. Left: cross-validated (5×5) Gradient Boosting performance across metrics using different sets of classical and quantum-extracted features, as well as their SHAP-based combination. Right: SHAP importance of the features selected as most relevant by the tree explainer.}
  \label{fig:scores_and_shap}
\end{figure*}

The results in Fig.~\ref{fig:scores_and_shap} show that the models learn effectively from quantum-extracted features and that these features yield clear gains. The SHAP analysis further quantifies the contribution of each feature subset.

For the molecular toxicity case, features extracted from two-body correlations provide the best standalone performance, surpassing the purely classical dataset. The concatenated data produce a substantial enhancement across all metrics compared with the standalone datasets, notably $121\%$ in precision. The feature-importance analysis shows that, among the 50 selected variables, 43 are quantum-extracted, accounting for $86.0\%$ of the explainer’s total importance. Notably, the two-body correlations rank as the most important ($40.0\%$), consistent with their best standalone performance. This proves that different quantum measurements from the same ansatz can be combined to improve model learning.

In the breast tumor detection problem, where features are extracted from $224\times 224$ images, our quantum feature extraction is applied on top of pre-extracted features, a sequential classical–quantum feature extraction. While we expect the strongest feature extractors to have access to raw data (and thus anticipate some loss of potential when operating on pre-extracted features), applying our quantum procedure still improves over purely classical models. The mean AUC shows a $5.5\%$ enhancement relative to the classically extracted dataset. Of the 156 selected variables, 109 are quantum-derived, totaling $44.4\%$ of the overall importance. On the MedMNIST predefined test set, the AUC is $0.937$, outperforming ResNet18 ($0.891$), ResNet50 ($0.866$), and Google Vision ($0.919$) whose parameter counts are $> 11\times 10^6$.

\begin{table}[h!]
\centering
\renewcommand{\arraystretch}{1.2} 
\setlength{\tabcolsep}{10pt}   
\begin{tabular}{lcc}
\hline\hline
\textbf{Method} & \textbf{AUC} & \textbf{Accuracy} \\
\hline
SVC (SHAP-selected)     & \textbf{0.937} & \textbf{0.876} \\
SVC (Original features) & 0.887 & 0.830 \\
GB (SHAP-selected)      & 0.919 & 0.827 \\
GB (Original features)  & 0.882 & 0.830 \\
\hline
ResNet-18          & 0.891 & 0.833 \\
ResNet-50          & 0.866 & 0.842 \\
Auto-sklearn            & 0.836 & 0.803 \\
AutoKeras               & 0.871 & 0.831 \\
Google AutoML Vision    & 0.919 & 0.861 \\
\hline\hline
\end{tabular}
\vspace{2mm}
\caption{Test performance on the \textit{Breast MedMNIST} dataset. Comparison between traditional machine learning models (SVC, GB) trained with SHAP-based feature selection and the state-of-the-art deep learning methods.}
\label{tab:breast_test}
\end{table}

\section{Conclusion}
\label{sec:conclusion}

We have demonstrated a scalable quantum feature extraction framework that leverages $k$-body Hamiltonian encoding and non-adiabatic quantum dynamics to map classical data into enriched feature spaces. By embedding both single-variable information and higher-order statistical correlations within spin-glass Hamiltonians and evolving them under counterdiabatic dynamics in the impulse regime, the proposed method generates expressive, nonlinearly correlated features that enhance classical machine learning performance.

Our implementation employed IBM quantum devices with up to 156 physical qubits (IBM \textit{Kingston}) to simulate the encoding Hamiltonians corresponding to both molecular toxicity and medical image classification tasks. Despite hardware connectivity constraints, the method preserved scalability and achieved statistically significant performance gains. In particular, the combined classical–quantum feature sets improved precision by $121\%$ in molecular toxicity classification, and enhanced the mean AUC by $5.5\%$ for breast tumor detection compared to purely classical models. On the \textit{Breast MedMNIST} benchmark, the hybrid approach achieved an AUC of $0.937$, outperforming deep-learning baselines such as ResNet18 ($0.891$) and ResNet50 ($0.866$), despite operating with only 156 features.

The feature-importance analysis using SHAP revealed that the quantum-derived observables contributed dominantly to model decisions. For example, in the molecular toxicity problem, $86\%$ of the total SHAP importance originated from quantum features, indicating that the generated observables encode distinct and complementary information not captured by classical preprocessing. This strong SHAP-weighted contribution serves as an empirical signature of the quantumness of the data, that is, correlations and nonlinear structures induced by quantum evolution that improve generalization in classical models.

Overall, these results highlight a practical and hardware-efficient approach to building hybrid classical–quantum pipelines, where quantum dynamics act as a physically grounded mechanism for generating rich, nonlinear features. The consistent improvements observed across real-world datasets, together with the clear dominance of quantum-derived features in SHAP-based interpretability analyses, demonstrate that the proposed feature extraction procedure introduces genuine, data-relevant structure beyond classical reach. This framework, therefore, outlines a realistic and near-term pathway for quantum-assisted learning, where quantum dynamics complement classical models to deliver measurable performance gains in data-driven applications.

\appendix
\twocolumngrid 
\onecolumngrid 

\section{Variable--to--Qubit Assignment Optimization}\label{app:a}
We consider a quantum hardware graph corresponding to the \textit{IBM Kingston} device, which contains 156 qubits. This hardware graph defines the allowed connectivity between qubits, meaning that only certain pairs $(i,j)$ can support two-body interactions, and specific triplets $(i,j,k)$ can support three-body interactions.  

Given a classical dataset with exactly 156 features (one per qubit), our objective is to determine an optimal mapping of features onto qubits so as to maximize the amount of statistical dependency preserved within the hardware’s limited connectivity. To achieve this, we define the interaction coefficients $J_{ij}$ and $J_{ijk}$, which quantify the normalized mutual information between feature pairs and the normalized three-way mutual information among feature triplets, respectively. In practice, the three-body term $J_{ijk}$ is approximated using the average of the corresponding pairwise mutual information values.  

Only those interactions within the Marrakesh hardware connectivity graph are considered valid, corresponding to the sets of edges $E$ and triplets $T$. The optimization process is carried out using a Genetic Algorithm (GA), which searches for a permutation $\pi$ that assigns each feature to a specific qubit. The fitness function is defined as  

\begin{equation}
    F(\pi) = \lambda_2 \sum_{(i,j)\in E} J_{\pi(i),\pi(j)} \;+\; 
         \lambda_3 \sum_{(i,j,k)\in T} J_{\pi(i),\pi(j),\pi(k)}
\end{equation}
where the parameters $\lambda_2$ and $\lambda_3$ regulate the relative influence of the two-body and three-body interaction terms.  

The Genetic Algorithm employs tournament selection, crossover, and mutation operators to evolve candidate assignments, producing an optimized variable placement that maximizes information preservation under the Marrakesh hardware connectivity.

\vspace{0.5cm}
\noindent\textbf{Pseudocode:}

\begin{algorithm}[H]
\SetAlgoLined
\SetAlgoSkip{medskip}    
\DontPrintSemicolon      
\SetKwInput{KwIn}{Input}
\SetKwInput{KwOut}{Output}

\KwIn{Dataset $X$ with $N=156$ features, hardware graph $G=(E,T)$, GA hyperparameters, weights $\lambda_2, \lambda_3$}
\KwOut{Optimized mapping $\pi^\ast$ of features to qubits}

\BlankLine

\textbf{Step 1: Compute interaction matrices}\;
\For{all feature pairs $(x_i,x_j)$}{
    $J_{ij} \gets \text{normalized mutual information}(x_i, x_j)$
}
\For{all triplets $(x_i,x_j,x_k)$}{
    $J_{ijk} \gets \text{normalized 3-way mutual information}(x_i, x_j, x_k)$
}

\BlankLine

\textbf{Step 2: Define fitness function}\;
Given assignment $\pi$:\;
\Indp
$F(\pi) = \lambda_2 \sum_{(i,j)\in E} J_{\pi(i),\pi(j)} + 
          \lambda_3 \sum_{(i,j,k)\in T} J_{\pi(i),\pi(j),\pi(k)}$
\Indm

\BlankLine

\textbf{Step 3: Initialize GA population}\;
$P \gets$ random permutations of length $N$

\BlankLine

\textbf{Step 4: Evolution loop}\;
\For{$g = 1$ \KwTo $N_{gen}$}{
    Evaluate fitness of each individual in $P$\;
    Select parents via tournament selection\;
    Apply crossover to generate offspring\;
    Apply mutation with given probability\;
    $P \gets$ offspring
}

\BlankLine

\textbf{Step 5: Return best solution}\;
$\pi^\ast \gets \arg\max_{\pi \in P} F(\pi)$\;
$F^\ast \gets \max_{\pi \in P} F(\pi)$\;

\Return{$\pi^\ast, F^\ast$}

\caption{Genetic Algorithm for Variable--to--Qubit Assignment}
\end{algorithm}


\begin{figure*}[!htbp] 
  \centering
\includegraphics[width=\textwidth,height=\textheight,keepaspectratio]{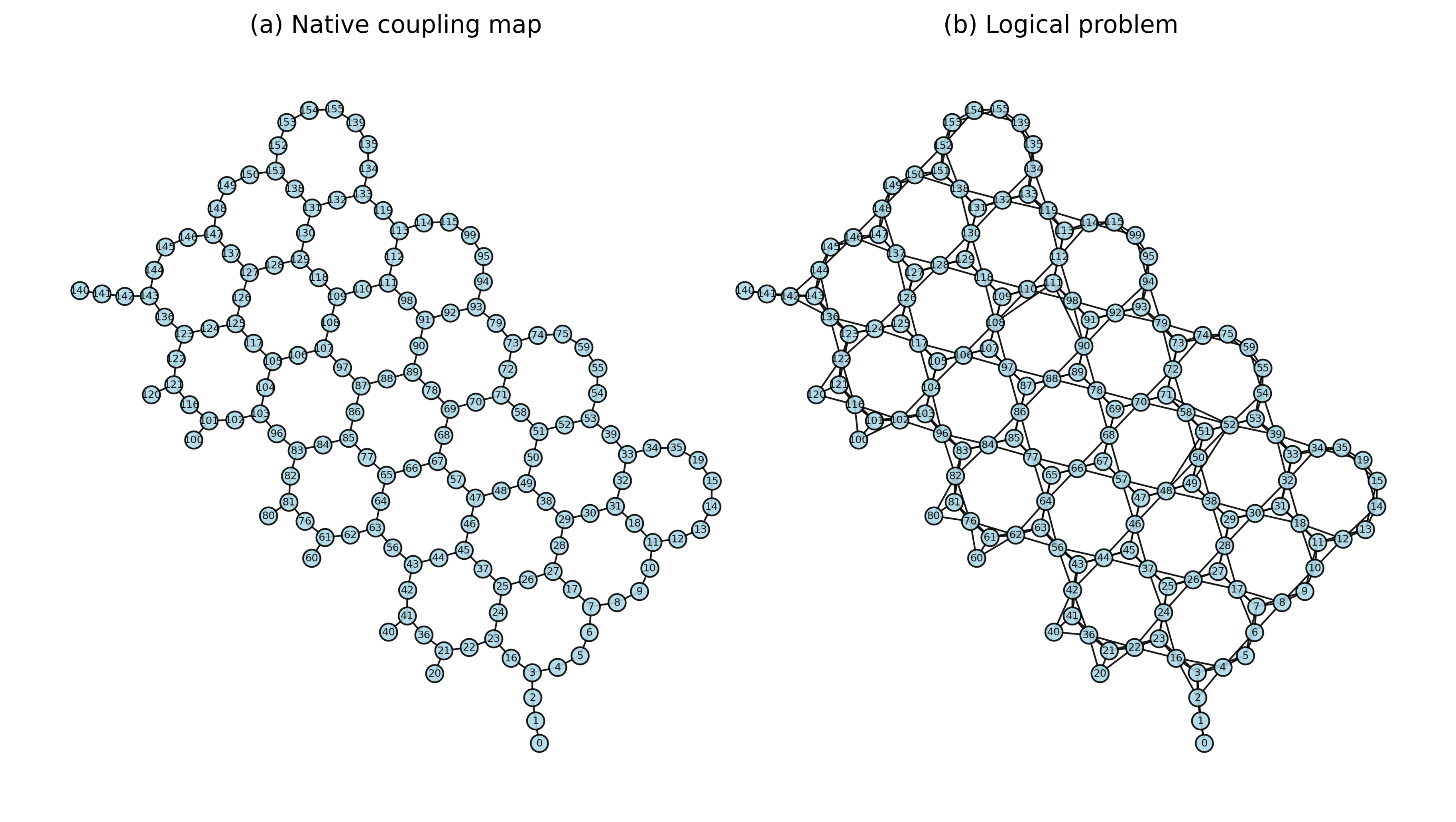}
  \caption{Illustration of the (b) embedded (logical) problem compared to the heavy-hexagonal topology extracted from the  (a) native connections of IBM Kingston.}
  \label{fig:embedding_graphs}
\end{figure*}

\begin{figure}[!htbp]
  \centering
  \includegraphics[width=0.7\textwidth,keepaspectratio]{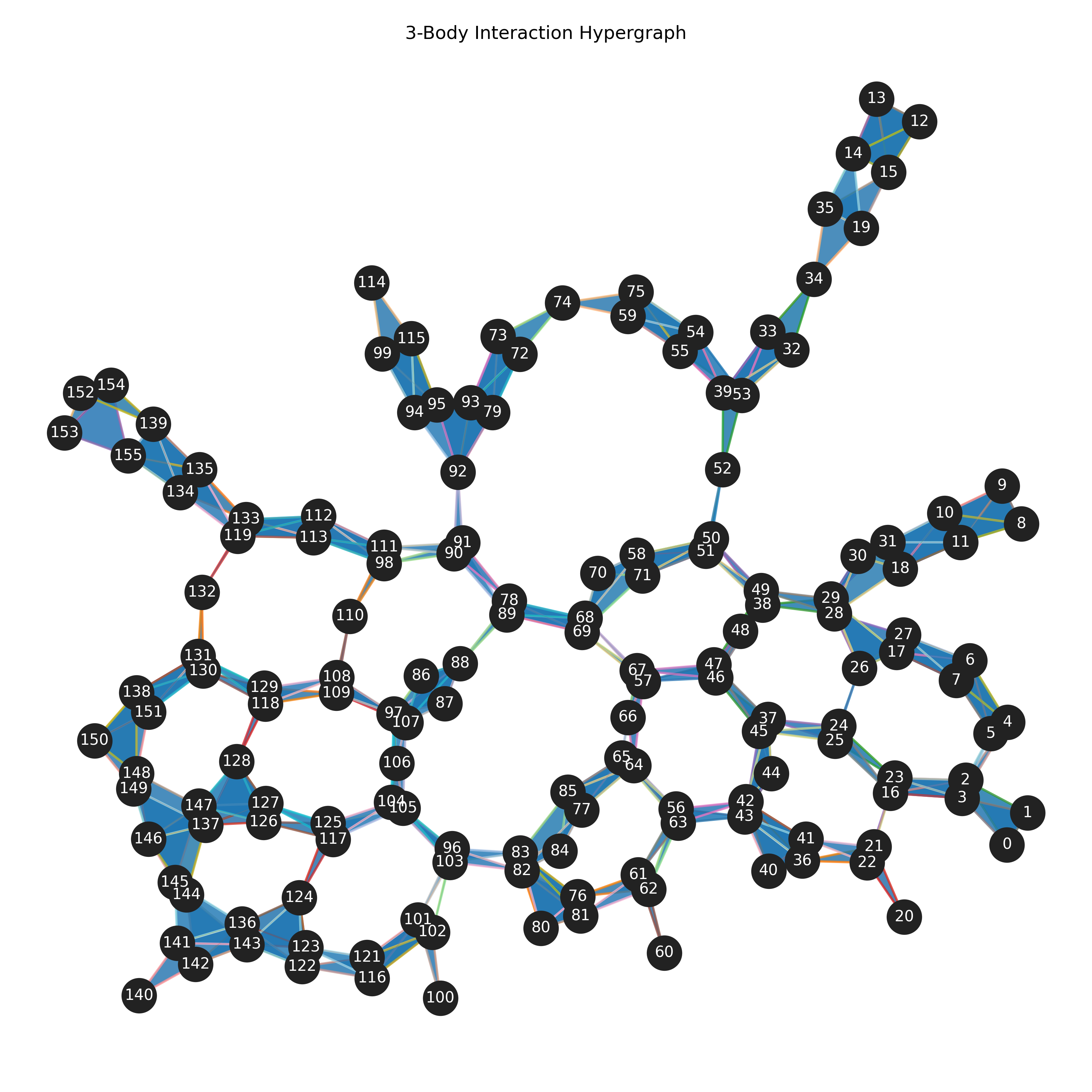}
  \caption{
  Visualization of a 3-uniform hypergraph representing all three-body interactions among spin variables. 
  Each triangular hyperedge corresponds to a third-order term in the encoded spin-glass Hamiltonian, where the presence and strength of a hyperedge encode the magnitude of the associated third-order mutual information between the connected variables. 
  This hypergraph structure forms the basis for constructing higher-order energy landscapes that capture nonlinear correlations beyond pairwise couplings.
  }
  \label{fig:hypergrah}
\end{figure}

The relation between qubit assignment, Fig.~\ref{fig:embedding_graphs}, and  Fig.~\ref{fig:hypergrah} can be understood as follows. 
Fig.~\ref{fig:embedding_graphs} illustrates the hardware topology of the \textit{IBM Kingston} device, where the limited connectivity imposes strict physical constraints on the number of possible 2-body and 3-body interactions among qubits. 
While a fully connected system with 156 qubits would allow for more than $6\times10^5$ potential triplets, the actual hardware graph admits only 222 valid 3-qubit interactions. 
This sparse connectivity means that an arbitrary or sequential feature-to-qubit assignment would disregard many statistically relevant relations (high $J_{ijk}$ values), leading to a suboptimal encoding of the classical correlations within the quantum system.  

To overcome this limitation, the Genetic Algorithm (GA) described above searches for an optimal permutation of features that maximizes the preservation of information across the available hardware links. 
The optimization objective follows the general formulation in Eq.~(A1), where the term $\sum J_{ij}$ is recovered when only two-body connections are considered ($\lambda_3=0$), and $\sum J_{ijk}$ becomes dominant when including three-body interactions ($\lambda_2=0$). 
The underlying hypergraph shown in Fig.~\ref{fig:hypergrah} captures these higher-order dependencies through its triangular edges, each representing a third-order coupling term. 

\section{Higher-order mutual information}\label{app:c}
The coefficients of the $k$-body terms in the encoding Hamiltonian (Eq.~\ref{eq:encoding_hamiltonian}) are given by the higher-order mutual information, defined as
\begin{equation}
    c_{S} = \frac{1}{\binom{m}{2}} 
            \sum_{\{a,b\}\subset S} I(x_a, x_b),
\end{equation}
where $I(x_a, x_b)$ represents the pair-wise mutual information between two random variables $x_a$ and $x_b$ from the dataset. 

Mutual Information (MI) quantifies the statistical dependency between two random variables by measuring the reduction in uncertainty of one variable given knowledge of the other. 
Unlike linear correlation coefficients such as Pearson’s $r$, MI is rooted in information theory and depends on probability distributions rather than linear relationships. 
It captures both linear and nonlinear dependencies by evaluating how much the joint probability distribution $p(x_a, x_b)$ diverges from the product of the marginals $p(x_a)p(x_b)$:
\[
I(x_a, x_b) = \sum_{x_a,x_b} p(x_a,x_b)\,\log\frac{p(x_a,x_b)}{p(x_a)p(x_b)}.
\]
A value of $I(x_a, x_b) = 0$ indicates complete statistical independence, while higher values denote stronger dependence, regardless of the underlying functional form.

In the context of our encoding Hamiltonian, the use of MI provides a natural way to quantify the shared information between features in the classical dataset before embedding them into the quantum system. 
By averaging the pairwise MI values among all variables within a subset $S$, the higher-order coefficient $c_S$ summarizes the collective dependency structure of the variables that participate in a $k$-body interaction. 
This formulation allows the encoded Hamiltonian to incorporate complex nonlinear and synergistic relationships among variables that would otherwise be missed by purely linear measures, thus producing a more expressive and information-preserving quantum representation.

\section{Feature extraction comparison with classical methods}\label{app:d}

To assess the relative importance of the quantum-derived features compared to classical feature extraction techniques, we performed a SHAP-based analysis of feature relevance using a Gradient Boosting model. 
For both the \textit{Molecular Toxicity} and \textit{BreastMNIST} datasets, we evaluated the top 100 most relevant variables among the different feature groups: local magnetizations (1-body), two-body correlations, three-body correlations, PCA components, UMAP embeddings, and the original input features.

\begin{figure}[h!]
    \centering
    \makebox[\textwidth][c]{\textbf{(a)}\hspace{0.5cm}}
    \includegraphics[width=0.6\textwidth]{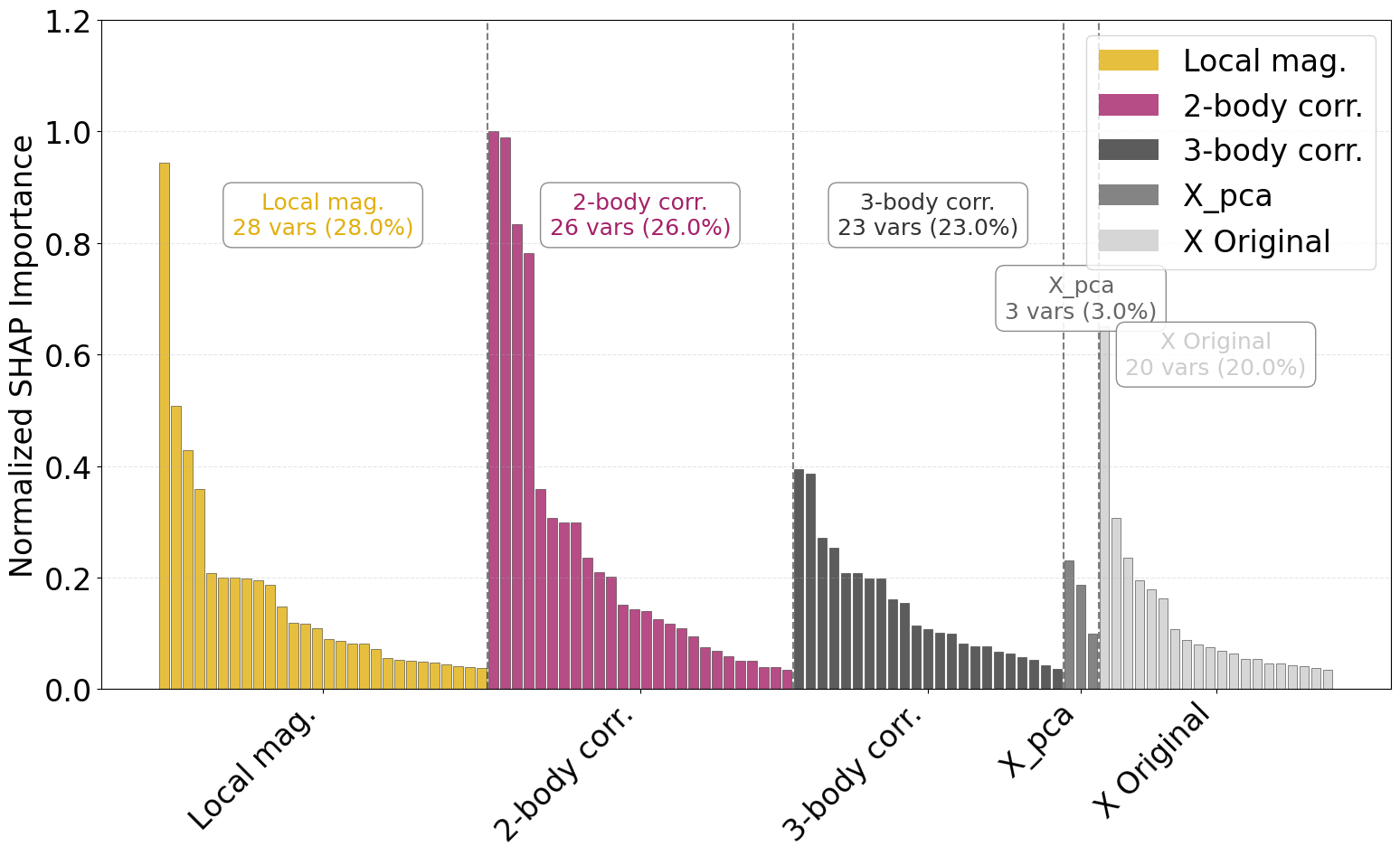}\\[2mm]
    \makebox[\textwidth][c]{\textbf{(b)}\hspace{0.5cm}}
    \includegraphics[width=0.6\textwidth]{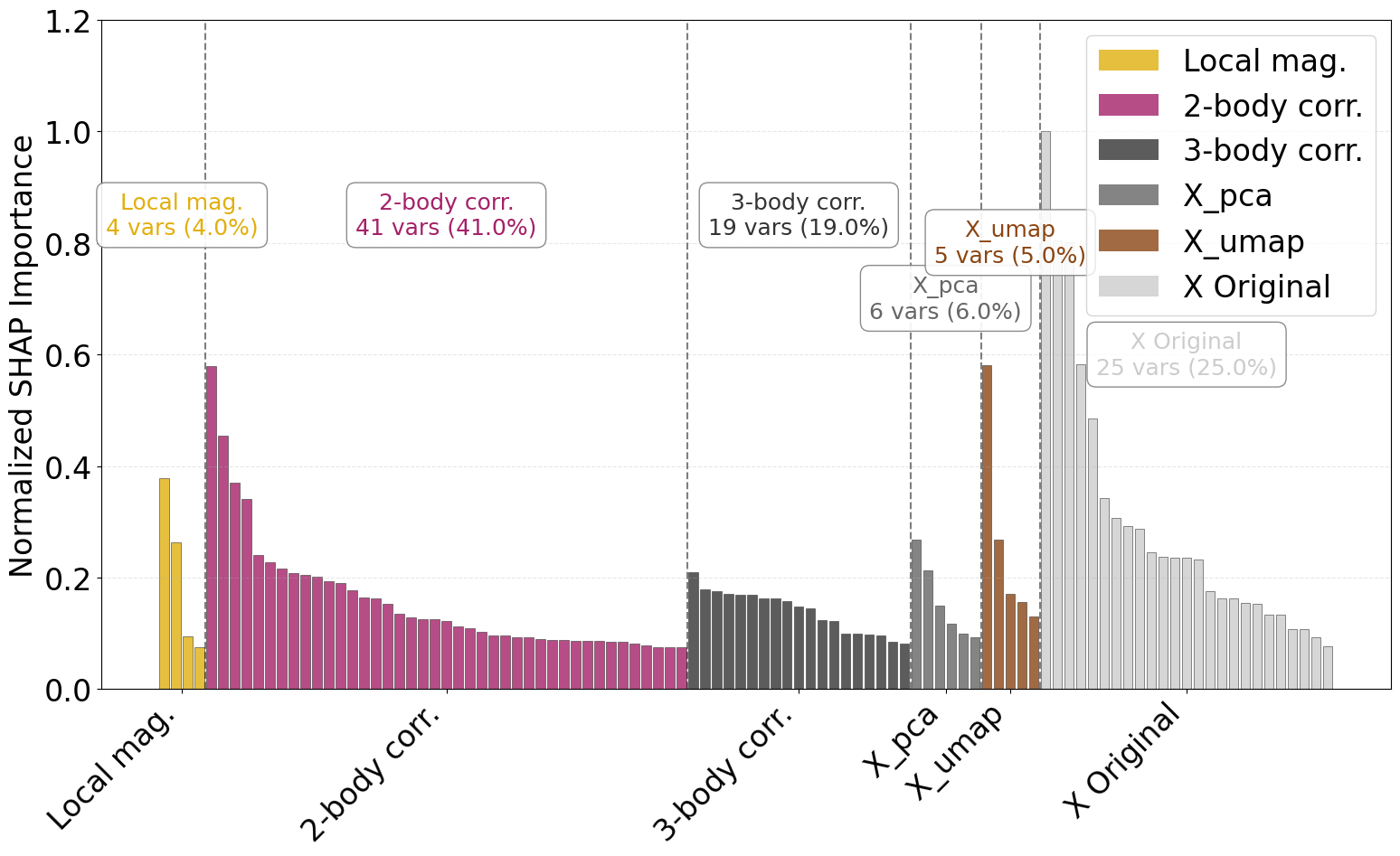}
    \vspace{2mm}
    \caption{Normalized SHAP importance of the 100 most relevant features obtained using a Gradient Boosting model. 
    (a)~\textit{Molecular Toxicity} dataset. Only three PCA features appear among the top 100, while no UMAP features are selected. 
    (b)~\textit{BreastMNIST} dataset. The two-body correlation features show the highest relevance, followed by UMAP and three-body correlations.}
    \label{fig:feature_importance_comparison}
\end{figure}

Fig.~\ref{fig:feature_importance_comparison} shows the normalized SHAP importance distributions for each feature category. 
In the case of the molecular toxicity dataset [Fig.~\ref{fig:feature_importance_comparison}(a)], only three PCA components appear among the top 100 variables, while no UMAP-derived features are selected, indicating their limited contribution to model interpretability. 
In contrast, the quantum-inspired features, especially the two-body correlation terms, exhibit consistently higher SHAP importance scores, suggesting that they capture a richer and more relevant representation of the underlying data structure.

For the \textit{BreastMNIST} dataset [Fig.~\ref{fig:feature_importance_comparison}(b)], a similar overall trend is observed, with the two-body correlation features again showing the highest relevance according to the SHAP analysis. 
Interestingly, in this case several UMAP-derived components gain noticeable importance, surpassing the average contribution of the one-body and three-body terms, although they remain fewer in number. 
This indicates that manifold-based embeddings such as UMAP can capture certain nonlinear structures present in the data; however, the quantum-derived correlations still dominate the overall relevance distribution. 
Overall, PCA continues to have a minor impact compared to both quantum features and UMAP.  
These results confirm that quantum feature extraction provides a more expressive representation, with two-body correlations consistently emerging as the most discriminative features across datasets.

\clearpage
\twocolumngrid
\bibliography{bibfile}

\end{document}